\documentclass[12pt]{article}

\usepackage{amsmath,amssymb,cite,comment}

\def\cd{D}
\def\xiL#1#2{\xi^{1\,\cdots\,#1}_{1\,\cdots\,#2}\,}
\def\q{{\bf q}}

\def\hH{{\widehat H}}

\def\cp{{\cal P}} \def\cq{{\cal Q}}
\def\nn{\nonumber}

\font\verysmall=cmr5
\def\phr{\raise1.1pt\hbox{\verysmall x}\kern-8pt\supset}
\def\bhr{\raise1.1pt\hbox{\verysmall x}\kern-9pt\subset}

\def\bbz{\mathbb{Z}}
\def\bbc{\mathbb{C}}

  \def\g{\gamma} \def\d{\delta}

\def\xiL#1#2{\xi^{1\,\cdots\,#1}_{1\,\cdots\,#2}\,}
\def\hp{\hat{\varphi}}  \def\l{\lambda} \def\ca{{\cal A}}
\def\hca{\hat{\cal A}}

\def\cc{{\cal C}} \def\ce{{\cal E}} 
\def\cu{{\cal U}}  \def\cp{{\cal P}}
    
\def\cz{{\cal Z}}  
 \def\cy{{\cal Y}}
\def\r{\rho} \def\eps{\epsilon}

 \def\x{\xi} \def\z{\zeta}
  \def\ve{\varepsilon}
\def\vf{\varphi} \def\y{\eta} 
\def\tv{{\tilde\vf}} \def\tcc{{\hat \cc}}
  \def\tc{\tilde c}

  \def\br{{\bar r}} 
  
\def\be{{\bar \ce}}

\def\tB{{\tilde B}}

\def\be{\begin{equation}}
\def\eqn#1{\be\label{#1}}
\def\ee{\end{equation}}

\def\bea{\begin{eqnarray}}
\def\eqnn#1{\bea\label{#1}}
\def\eea{\end{eqnarray}}

\newcommand{\eqna}[1]{\begin{subequations} \label{#1}
\begin{eqnarray}}
\def\eena{\end{eqnarray}
\end{subequations}}

\def\nd{\vfill\end{document}}

\begin{document}

\begin{center}

{\LARGE {\bf Representations of Multiparameter Quantum Groups}}

\vspace{10mm}

{\bf \large V.K. Dobrev}

\vspace{5mm}

 Institute for Nuclear
Research and Nuclear Energy \\
{Bulgarian Academy of Sciences}\\ {72 Tsarigradsko Chaussee, 1784
Sofia, Bulgaria}

\end{center}
 \vspace{10mm}

 \begin{abstract}
 We  construct representations  of the quantum
algebras ~$U_{q{\bf q}}(gl(n))$\\ and ~$U_{q{\bf q}}(sl(n))$~ which
are in duality with the multiparameter quantum  groups
~$GL_{q\q}(n)$, ~$SL_{q\q}(n)$,~ respectively. These objects
  depend on ~$n(n-1)/2+1$~ deformation parameters ~$q,q_{ij}$ ($1\leq i <j\leq n$)
  which is the maximal possible number  in the
  case of $GL(n)$. The representations are
labelled by $n-1$ complex numbers ~$r_i$~ and are
acting in the space of formal power series of ~$n(n-1)/2$~
non-commuting variables.  These variables generate  quantum flag
manifolds  of ~$GL_{q\q}(n)$, ~$SL_{q\q}(n)$.
The case $n=3$ is treated in more detail.
\end{abstract}

\vspace{10mm}

 \section{Introduction}

About 30 years passed since the advent of quantum groups at
center-stage of modern mathematical physics, cf., e.g.,
\cite{Dra,Jia,Lua,Mab,Wor,Bie,DKa,FRT,Maj,KSb}. Yet the field is
growing stronger every day, cf. a recent review in \cite{Dogru}.
With the present paper we contribute to the continuing in-depth
studies of quantum groups by constructing representations of ~{\it
multiparameter} quantum groups on the example of such multiparameter
deformations of the group ~$GL(n)$.

We follow the approach of \cite{Dob} as adopted to quantum groups in
\cite{Dopq,Dof} (see also \cite{Dogru}). We start with the
multiparameter quantum  groups ~$GL_{q\q}(n)$, introduced by Sudbery
\cite{Sud} depending on ~$n(n-1)/2+1$~ deformation parameters
~$q,q_{ij}$ ($1\leq i <j\leq n$)   which is the maximal possible
number  in the  case of $GL(n)$. We  construct representations  of
the quantum algebras ~$U_{q{\bf q}}(gl(n))$\\ and ~$U_{q{\bf
q}}(sl(n))$~ (found in \cite{DoPar}) which are in duality with the
multiparameter quantum groups ~$GL_{q\q}(n)$, ~$SL_{q\q}(n)$,~
respectively.     The representations are labelled by $n-1$ complex
numbers ~$r_i$~ and are acting in the space of formal power series
of ~$n(n-1)/2$~ non-commuting variables.  These variables generate
quantum flag manifolds  of ~$GL_{q\q}(n)$, ~$SL_{q\q}(n)$. We treat
the case $n=3$ in greater detail.

\vskip 5mm

\section{Multiparametric deformation of GL(n)}
\label{MDGL}

Here we use  the quantum group deformation of ~$GL(n)$~ introduced
by Sudbery  \cite{Sud}. That deformation depends on the   maximal possible number of
parameters :  ~$N = n(n-1)/2 + 1$. We denote these $N$
parameters by $q$ and $q_{ij}$, $1 \leq i < j \leq n$, and also for
shortness by the pair $q,\q$.  The standard one-parameter deformation is obtained
by setting ~$q_{ij}=q$, $\forall ~i,j$.

Explicitly, we consider an ~$n\times n$~ quantum matrix ~$M$~ with
non-commuting matrix elements ~$a_{ij}$, $1 \leq i,j \leq n$.
The matrix quantum group ~$\ca \equiv GL_{q\q}(n)$~ is generated
by the matrix elements ~$a_{ij}$~ with the following
commutation relations \cite{Sud}: \eqna{su1}
 a_{ij} a_{i\ell} ~&=&~ p a_{i\ell} a_{ij} ~, ~{\rm
for} ~ j < \ell ~, \\ 
a_{ij} a_{kj} ~&=&~ r a_{kj} a_{ij} ~, ~{\rm for} ~ i < k ~, \\ 
p a_{i\ell} a_{kj} ~&=&~ r a_{kj} a_{i\ell} ~, ~{\rm for} ~ i < k
~, ~j < \ell ~, \\ 
r q a_{k\ell} a_{ij} &-& (qp)^{-1} a_{ij} a_{k\ell} ~=~ \l
a_{i\ell} a_{kj} ~, ~{\rm for} ~ i < k ~, ~j < \ell ~, \\ 
p &=& q_{j\ell} / q^2 ~, ~~ r = 1 / q_{ik} ~, ~~\l = q -
1/q ~. \eena 
Considered as a bialgebra, it has the  standard
comultiplication ~$\d_\ca$~ and
counit ~$\ve_\ca$~:
\eqn{su2} \d_\ca(a_{ij}) ~=~ \sum_{k=1}^n a_{ik} \otimes a_{kj} ~,
~~~\ve_\ca(a_{ij}) ~=~ \d_{ij} ~. \ee
For the antipode we refer to \cite{Sud}.

Following the approach of \cite{Dob} (see also
\cite{Doba,Dobb,Dobc}) we shall use representations of the dual
quantum algebra on suitable quantum flag manifolds of ~$\ca$. For
this we first use the triangular decomposition of $\ca$~ \cite{Dof}:
\eqn{gau} a_{i\ell} ~=~ \sum_j Y_{ij} D_{jj} Z_{j\ell} ~,\ee where
\eqnn{gaus}
Y_{ij} ~&=&~ \xiL{j-1\,i}{j     } \cd_{j}^{-1} ~,\\
Z_{j\ell} ~&=&~ \cd_i^{-1} \xiL{j     }{j-1\,\ell} ~,\\
D_{jj} ~&=&~ \cd_{j} \cd_{j -1}^{-1} ~,\\
\cd_m ~&=&~ \sum_{\r \in S_m} ~
\eps (\r) ~ a_{1,\r (1)} \ldots a_{m,\r (m)} \ , \\
\xi^I_J ~&=&~ \sum_{\r\in S_r} \eps(\r)~
a_{i_{\r(1)}j_1}\cdots a_{i_{\r(r)}j_r} ~, \\
I&=&\{ i_1 < \cdots < i_r \} ~, ~~~J=\{ j_1 < \cdots < j_r \} \ ,
\eea $S_n$ is the permutation group of $n$ elements. Note that
$Y_{i\ell}=0$ for $i<\ell$, ~$Y_{ii}=1_\ca$, ~$Z_{i\ell}=0$ for
$i>\ell$, ~$Z_{ii}=1_\ca$, ~$\cd_0 \equiv 1_\ca$, ~$\xiL{i  }{i  } =
\cd_i$~.   Then ~$\cy_{q\q} \equiv \{Y_{j\ell},\ j>\ell\}$, ~may be
regarded as a quantum analogue of  the flag manifold ~$GL(n)/DZ$,
~$\cz_{q\q} \equiv \{Z_{ij}, ~i<j\}$, ~may be regarded as a quantum
analogue of  the flag manifold ~$B\backslash GL(n)$.

We give the commutation relation between the generators ~
$\{Y_{ji}\}$~ since we shall build our representations on
~$\cy_{q\q}\,$. The indices used below obey ~$1\leq i<j<k<l\leq n$.
We also use the notation: \eqn{ppp} p_{ij} ~\equiv~ {q_{ij} \over
q^2} ~, \quad p'_{ij} ~\equiv~ {q'_{ij} \over q'^2 } ~, \quad q'
~\equiv~ 1/q ~, \quad q'_{ij}  ~\equiv~ q_{ij}/q^2 \ee We have:
\eqna{cyy} &&Y_{kj}Y_{ki} ~=~ {q_{ij}q_{jk} \over q_{ik}}
Y_{ki}Y_{kj}
\ , \\ 
&&Y_{ki}Y_{ji} ~=~ {q_{ij}q_{jk} \over q_{ik}} Y_{ji}Y_{ki}
\ , \\ 
&&Y_{kj}Y_{ji} ~=~ {p_{ij}p_{jk}\over p_{ik}}Y_{ji}Y_{kj} ~+~
u^{-1}(u-u^{-1})Y_{ki}  \ , \\ 
&&Y_{li}Y_{kj} ~=~{q_{ik}q_{kl}\over q_{ij}q_{jl}}
Y_{kj}Y_{li} \ , \\ 
&&{q_{jl}\over q_{jk}q_{kl}}Y_{lj}Y_{ki} ~=~
{p_{ij}p_{jl}\over p_{il}}Y_{ki}Y_{lj} ~+~
u^{-1}(u-u^{-1})Y_{kj}Y_{li} \ , \\ 
&&Y_{lk}Y_{ji} ~=~ {q_{ik}q_{jl}\over q_{il}q_{jk}}Y_{ji}Y_{lk}
\eena 

\section{Multiparameter dual algebra}

In \cite{DoPar} we have found the dual to ~$\ca$~ algebra ~$\cu_g
\equiv U_{q{\bf q}}(gl(n))$. We fix the standard decomposition
~$gl(n) = sl(n) \oplus \cz$, where ~$\cz$ is the central subalgebra
of ~$gl(n)$.

The Drinfeld--Jimbo form of the dual commutation algebra
~$\cu_g$~ in terms of the $sl(n)$ generators ~$H_i,X^\pm_i$~ and
the $\cz$ generator $K$ is given as follows:
\eqna{wb}
 [ H_i , X^\pm_j ] ~&=&~ \pm c_{ij} X^+_j\, \\ 
{} [X^+_i , X^-_i ] ~&=&~ \l^{-1} \left( q^{H_i} - q^{-H_i} \right) ~\equiv~ [ H_i]_q \\
{} [K,Y]~&=&~ 0, \qquad \forall Y \in sl(n)
\eena 
where ~$\l \equiv q-q^{-1}$, ~$c_{ij}$~ is the standard Cartan
matrix of ~$gl(n,\bbc)$.

Thus as a ~{\it commutation algebra}~ we have the splitting ~ $\cu_{q{\bf q}} \cong
U_q(sl(n,\bbc)) \otimes U_q(\cz)$, and dependence only on the parameter ~$q$.

 This splitting is preserved also by the co-unit and the antipode:
\eqna{wh}
\ve_\cu (Y) ~&=&~ 0 ~, \quad Y=X^\pm_i\ ,H_i\ ,K,  \\ 
 \g_\cu (X^\pm_i) ~&=&~    -q^{\pm 1}\, (X^\pm_i)\ , \quad
 \g_\cu (Y) ~=~ -Y \ , ~~ Y=H_i,K , \eena
 and by the coproducts of $H_i,K$~:
\eqn{whh}\d_\cu (Y) ~=~ Y \otimes 1_\cu ~+~ 1_\cu \otimes Y,
\quad Y=H_i,K \ee

However, for the coproducts of the Chevalley generators $X^\pm_i$
we have:
\eqna{wg}
  \d_\cu (X^+_i) ~&=& ~ X^+_i \otimes \cp^{1/2}_i ~+~
\cp^{-1/2}_i \otimes X^+_i ~, \\ 
\d_\cu (X^-_i) ~&=& ~ X^-_i \otimes \cq^{1/2}_i ~+~
\cq^{-1/2}_i \otimes X^-_i ~, \eena 
where:
\eqnn{uu} \cp_i ~&=&~  \left( \prod_{s=1}^{i-1}
\left( {q_{si} \over q_{s,i+1}}\right)^{\hH_s} \right)
\left( {q^2 \over q_{i,i+1} }\right)^{\hH_i}
\left( {1 \over q_{i,i+1}} \right)^{\hH_{i+1}} \
\prod_{t=i+2}^{n-1} \left( {q_{i+1,t} \over q_{it}}
\right)^{\hH_t} \ ,  \nn\\
\cq_i ~&=&~ q^{2H_i}\cp_i\ ~, \quad \hH_i ~\equiv ~ \sum_{j=i}^{n-1} H_j
 \eea

Thus, the coproduct structure is not split and depends on all
parameters. Yet for a special choice of ~$n-1$~ of the parameters
(e.g., $q_{i,i+1}$) ~$\cu_g$~ can be split as a direct product of
two Hopf subalgebras: ~$\cu \equiv U_{q{\bf q}}(sl(n))$~ and
~$U_q(\cz)$, where ~$\cu$~ depends only on ~$(n^2 -3n+4)/2$~
parameters \cite{DoPar}.

Further we shall work with ~$\cu$~ in the setting of this complete
splitting.

\section{Representations  of the dual algebra}

We shall work with representation spaces of ~$\cu$~ parametrized by $n-1$ numbers ~$r_i$~ which will
be integers initially. The elements of these  spaces
 will be formal power series:
\eqnn{serasa}
\tv ({\bar Y}, {\bar \cd})
~&=&~ \sum_{{ \bar m \in\bbz_+}}
\mu_{\bar m} ~
(Y_{21})^{m_{21}} \ldots (Y_{n,n-1})^{m_{n,n-1}}
(\cd_{1})^{r_{1}} \ldots (\cd_{n-1})^{r_{n-1}} ~= \nn\\ 
~&=& ~~\hp ({\bar Y})~
(\cd_{1})^{r_{1}} \ldots (\cd_{n-1})^{r_{n-1}} ~, \eea 
where ${\bar Y}, {\bar \cd}$ denote the variables $Y_{il}$, $i>\ell$,
$\cd_i$, $i<n$.

First we shall give the left representation
action ~$\pi$~ of ~$\cu$~  on ~$\hp$. Besides the action of the 'Chevalley'
generators ~$K_i\equiv q^{H_i},X^\pm_i$~ we shall give for the
readers convenience also the action of ~$\cp_i,\cq_i$~
though it follows from that of ~$K_i$. We have:
\eqna{ulay}
 \pi(K_i)Y_{lj}~&=&~q^{(\d_{i+1,l}-\d_{i+1,j}
-\d_{il}+\d_{ij})/2}Y_{lj} \ , \\ 
\pi(X^+_i)Y_{lj}~&=&~-q\,Q_{i,i+1}^{-1/2}
Q_{ij}^{-1/2}\d_{il}Y_{l+1,j}) ~+ \nn\\
&&+~q\,Q_{i,i+1}^{-1/2}Q_{il}^{-1/2}
\left({q_{j,j+1}q_{j+1,l}\over q_{jl}}\right)^{(1-\d_{l,j+1})}
\d_{ij}Y_{j+1,j}Y_{lj}~+ \nn\\
&&+~q\,Q_{i,i+1}^{-1/2}Q_{il}^{1/2}Q_{i,j-1}^{-1/2}
Q_{ij}^{-1/2}\d_{i+1,j}~\times\nn\\
&&\times~\{{q_{j-1,l}\over q_{j-1,j}q_{jl}}Y_{l,j-1}
-Y_{j,j-1}Y_{lj}\}  \ , \\ 
\pi(X^-_i)Y_{lj}~&=&~-q^{-2}Q_{ii}^{1/2}Q_{ij}^{1/2}q^{-\d_{ij}}\d_{i+1,l}
Y_{l-1,j} \ , \\ 
\pi(\cp_i^{1/2})Y_{lj}~&=&~Q_{il}^{-1/2}Q_{ij}^{1/2}Y_{lj} \ , \\ 
\pi(\cq_i^{1/2})Y_{lj}~&=&~q^{(\d_{i+1,l}-\d_{i+1,j}
-\d_{il}+\d_{ij})}Q_{il}^{1/2}Q_{ij}^{-1/2}Y_{lj} \eena 
where
\eqn{pid}
Q_{is}~=~\begin{cases}{ q_{si}\over q_{s,i+1}}~, &  s\leq i-1 \cr
{q^2\over q_{i,i+1}} ~, &  s= i \cr
{1\over q_{i,i+1}}~, &  s= i+1 \cr
{q_{i+1,s}\over q_{is}}~, &  s\geq i+2 \end{cases} \ee  

The above is supplemented with the following action on the unit
element of $\ca$\ :
\eqn{lru}
\pi (K_i) ~1_\ca ~~=~~ 1_\ca ~, \quad  \pi (X^\pm_i) ~1_\ca ~~=~~
0 ~\ee

In order to derive the action of $\pi(y)$ on arbitrary elements of
the basis \eqref{serasa}, we use the twisted derivation rule consistent with the
coproduct and the representation structure, namely, we take:
~$\pi (y)\vf\psi ~=~ \pi (\d'_{\cu}(y))(\vf\otimes \psi)$,
~where ~$\d'_{\cu} ~=
~\sigma \circ \d_{\cu}$ ~is the opposite coproduct,
($\sigma$ is the permutation
operator). Thus, we have: \eqna{tw}
 \pi (K_i)\vf\psi ~&=&~ \pi (K_i)\vf\cdot \pi
(K_i)\psi\ , \\ 
\pi (X^+_i)\vf\psi ~&=&~ \pi (X^+_i)\vf\cdot \pi (\cp^{-1/2}_i)\psi
~+~ \pi (\cp^{1/2}_i)\vf\cdot \pi (X^+_i)\psi\  , \\
\pi (X^-_i)\vf\psi ~&=&~ \pi (X^-_i)\vf\cdot \pi (\cq^{-1/2}_i)\psi
~+~ \pi (\cq^{1/2}_i)\vf\cdot \pi (X^-_i)\psi\
\eena 

From now on we suppose that none of the deformation
parameters ~$q,q_{ij}$~ is  a nontrivial root of unity.

Applying \eqref{tw} to \eqref{ulay} we have:
\eqna{ulpy}
\pi(K_i)(Y_{lj})^k~ &=&~q^{k(\d_{i+1,l}-\d_{i+1,j}
-\d_{il}+\d_{ij})/2}(Y_{lj})^k \ , \\ 
\pi(X^+_i)(Y_{lj})^k~&=&~-q\,Q_{i,i+1}^{-1/2}
Q_{ij}^{(k-2)/2}c_l\d_{il}(Y_{lj})^{k-1}\
Y_{l+1,j}~+ \cr
&&+~q\,Q_{i,i+1}^{-1/2}Q_{il}^{(k-2)/2}c_j
\left({q_{j,j+1}q_{j+1,l}\over q_{jl}}\right)^{(1-\d_{l,j+1})}
\d_{ij}Y_{j+1,j}(Y_{lj})^k~+ \cr
&&+~q\,Q_{i,i+1}^{-1/2}Q_{il}^{k/2}({q_{j-1,j}\over q})^{k}\tc_{j-1}
\d_{i+1,j}~\times\cr
&&\times~\{{q_{j-1,l}\over q_{j-1,j}q_{jl}}Y_{l,j-1}(Y_{lj})^{k-1}
-Y_{j,j-1}(Y_{lj})^k\} \ , \\ 
\pi(X^-_i)(Y_{lj})^k~&=&~-q^{-2}Q_{ii}^{1/2}Q_{ij}^{k/2}q^{-k\d_{ij}}
c_{l-1}\d_{i+1,l}
Y_{l-1,j}(Y_{lj})^{k-1} \ , \\ 
\pi(\cp_i^{1/2})(Y_{lj})^k~&=&~Q_{il}^{-k/2}Q_{ij}^{k/2}(Y_{lj})^k
\ , \\ 
\pi(\cq_i^{1/2})(Y_{lj})^k~&=&~q^{k(\d_{i+1,l}-\d_{i+1,j}
-\d_{il}+\d_{ij})}Q_{il}^{k/2}Q_{ij}^{-k/2}(Y_{lj})^k
\eena 
\eqn{dcbc} c_i~=~ (q_{i,i+1})^{(k-1)/2}[k]_q, ~~
\tc_i~=~ (q_{i,i+1})^{(1-k)/2}[k]_q, \ee
$[k]_q=(q^k-q^{-k})/\l$.
\eqna{ulpd}
\pi(K_i)(D_j)^k~&=&~q^{-k\d_{ij}/2}(D_j)^k \ , \\ 
\pi(X^+_i)(D_j)^k~&=&~-q\,Q_{i,i+1}^{-1/2}
\left(\prod_{s=1}^{j-1}Q_{is}^{k/2}\right)\tc_j\d_{ij} \ Y_{j+1,j}(D_j)^k \ , \\ 
\pi(X^-_i)(D_j)^k~&=&~0  \ , \\ 
\pi(\cp_i^{1/2})(D_j)^k~&=&~\left(\prod_{s=1}^{j}Q_{is}^{-k/2}\right)(D_j)^k \ , \\ 
\pi(\cq_i^{1/2})(D_j)^k~&=&~q^{-k\d_{ij}}\left(\prod_{s=1}^{j}Q_{is}^{k/2}\right)(D_j)^k
\eena 

The action of $\cu$ on arbitrary elements $\tv$,  $\hp$ is found by combining the formulae
\eqref{ulpy},\eqref{ulpd} via \eqref{tw}.

\section{Case of multiparameter quantum GL(3)}

We  restrict now to case of $GL(3)$ which has a flag manifold ~$\cy
~=~ GL(3)/\tB ~=$ $=~ SL(3)/B$, where $\tB,B$ are the Borel subgroups of
$GL(3),SL(3)$, respectively, consisting of all upper diagonal
matrices.

  The multiparameter deformation
~$GL_{q{\bf q}}(3)$~   depends on ~$(n^2 -n+2)/2=4$~ parameters
~$q,q_{ij}$, ~$1\leq i<j\leq 3$.  Thus, the flag manifold
~$\cy_{q{\bf q}} ~=$ $~ GL_{q{\bf q}}(m)/\tB_{q{\bf q}}(m)$~ depends
on the same number of parameters. If we want to
 achieve the complete splitting of ~$\cu ~=~ U_{q{\bf q}}(sl(m))$~
we have to impose some relations between the parameters,
thus ~$\cu$~ would depend on ~$(n^2 -3n+4)/2=2$~ parameters.
  For this we have to impose that the parameters ~$q_{i,i+1}$~ are expressed through the
rest as: \eqn{split}
 q_{12} ~=~ q_{23} ~=~ {q^2 \over q_{13}}  \ . \ee 

 Further we note that in this case there are three
coordinates ~$Y_{ij}$~ of ~$\cy_{q\q}$. Their
  explicit commutation relations follow from \eqref{cyy} ($\l \equiv q-q^{-1}$):
  \eqnn{coo} &&Y_{32}Y_{31} ~=~ {q_{12} q_{23} \over
q_{13} }Y_{31}Y_{32} \ ,\\
 &&Y_{31}Y_{21} ~=~ {q_{12} q_{23} \over
q_{13}} Y_{21}Y_{31} \ ,\\
&&Y_{32}Y_{21} ~=~ {q_{12} q_{23} \over q^2 q_{13}} Y_{21}Y_{32} ~+~
q^{-1}\l  Y_{31} \nn\eea

Next we write down formulae \eqref{ulpy} in our case :
\eqna{ulpy3}
\pi(K_1)(Y_{lj})^k~ &=&~q^{k(\d_{2,l}-\d_{2,j}
+\d_{1j})/2}(Y_{lj})^k \ , \\
\pi(K_2)(Y_{lj})^k~ &=&~q^{k(\d_{3,l}
-\d_{2l}+\d_{2j})/2}(Y_{lj})^k \ ,\nn\\
\pi(X^+_1)(Y_{32})^k~&=&~
q_{12}\,q^{1-k} \left( {q_{12}q_{23}\over q_{13}}\right)^{k/2}   [k]_q
~\times\cr &&\times~
\{{q_{13}\over q_{12}q_{23}}Y_{31}(Y_{32})^{k-1}
-Y_{21}(Y_{32})^k\} \ , \\
\pi(X^+_1)(Y_{31})^k~&=&~
q\,q_{12} \left( {q_{12}q_{23}\over q_{13}}\right)^{k/2}[k]_q
 \,
Y_{21}(Y_{31})^k  \ , \nn\\
\pi(X^+_1)(Y_{21})^k~&=&~
q\,q_{12}[k]_q
 \ (Y_{21})^{k+1}  \ , \nn\\
\pi(X^+_2)(Y_{32})^k~&=&~-
q\,q_{23} [k]_q (Y_{32})^{k+1} \ , \\
\pi(X^+_2)(Y_{31})^k~&=&~ 0\ , \nn\\
\pi(X^+_2)(Y_{21})^k~&=&~-q\,q_{23}^{k/2}
\left( {q_{12}\over q_{13}}\right)^{(k-2)/2}   [k]_q (Y_{21})^{k-1}\ Y_{31}
  \ , \nn\\
\pi(X^-_1)(Y_{32})^k~&=&~ 0 , \\
\pi(X^-_1)(Y_{31})^k~&=&~ 0 , \nn\\
\pi(X^-_1)(Y_{21})^k~&=&~-q^{-1}
q_{12}^{-1} [k]_q (Y_{21})^{k-1} \ , \nn\\
\pi(X^-_2)(Y_{32})^k~&=&~-q^{-1}
 q_{23}^{-1} [k]_q (Y_{32})^{k-1} \ , \\
\pi(X^-_2)(Y_{31})^k~&=&~-q^{-1}
\left( {q_{12}q_{23}\over q_{13}}\right)^{k/2}
  [k]_q Y_{21}(Y_{31})^{k-1} \ , \nn\\
\pi(X^-_2)(Y_{21})^k~&=&~ 0, \nn\\
\pi(\cp_1^{1/2})(Y_{32})^k~&=&~ \left( {q_{13} \over q_{12}q_{23} }\right)^{k/2}\,
(Y_{32})^k ~=~ q^{-k}\ \pi(\cq_1^{-1/2})(Y_{32})^k\ , \\
\pi(\cp_1^{1/2})(Y_{31})^k~&=&~  \left( {q^2 q_{13} \over q_{12}q_{23} }\right)^{k/2}\,
(Y_{31})^k ~=~ q^k\ \pi(\cq_1^{-1/2})(Y_{31})^k  \ , \nn\\
\pi(\cp_1^{1/2})(Y_{21})^k~&=&~  q^k (Y_{21})^k  ~=~ \pi(\cq_1^{1/2})(Y_{21})^k
\ , \nn\\
\pi(\cp_2^{1/2})(Y_{32})^k~&=&~  q^k  \,(Y_{32})^k ~=~ \pi(\cq_2^{1/2})(Y_{32})^k\ , \\
\pi(\cp_2^{1/2})(Y_{31})^k~&=&~ \left( {q_{12}q_{23} \over q_{13}   }\right)^{k/2}\,
(Y_{31})^k ~=~ q^{k}\ \pi(\cq_2^{-1/2})(Y_{31})^k  \ , \nn\\
\pi(\cp_2^{1/2})(Y_{21})^k~&=&~ \left( {q_{12}q_{23} \over q^2 q_{13}   }\right)^{k/2}\,
(Y_{21})^k ~=~ q^{-k}\ \pi(\cq_2^{-1/2})(Y_{21})^k \
\eena
For the action on $D_j$ we take into account that their degrees in $\tv$ are fixed  as
 representation parameters:
\eqna{ulpd3}
\pi(K_i)(D_j)^{r_j}~&=&~q^{-r_j\d_{ij}/2}(D_j)^{r_j} \ , \\ 
\pi(X^+_i)(D_j)^{r_j}~&=&~-q\,Q_{i,i+1}^{-1/2}
\left(\prod_{s=1}^{j-1}Q_{is}^{r_j/2}\right)\tc_j\d_{ij} \ Y_{j+1,j}(D_j)^{r_j} \ , \\ 
\pi(X^-_i)(D_j)^{r_j}~&=&~0  \ , \\ 
\pi(\cp_i^{1/2})(D_j)^{r_j}~&=&~\left(\prod_{s=1}^{j}Q_{is}^{-r_j/2}\right)(D_j)^{r_j} \ , \\ 
\pi(\cq_i^{1/2})(D_j)^{r_j}~&=&~q^{-r_j\d_{ij}}\left(\prod_{s=1}^{j}Q_{is}^{r_j/2}\right)
(D_j)^{r_j}
\eena 

 Now we are ready to consider the representation action on the functions ~$\tv$~
from \eqref{serasa}  for which here we adopt the following notation:
 \eqnn{seras}
\tv ({\bar Y},{\bar D}) ~&=&~
\sum_{{ j,n,\ell \in\bbz_+}}\tv_{jn\ell} \ , \nn\\
\tv_{jn\ell} ~&=&~ Y_{21}^j \ Y_{31}^n \ Y_{32}^\ell\, D_1^{r_1} D_2^{r_2} \eea


Now the action of the representation $\pi$ is obtained using \eqref{ulpy3},\eqref{ulpd3} and
\eqref{tw}. Further we remove inessential phases. Finally we have:
\eqna{ulpiz}
\pi_{r_1,r_2}(K_1)\,\tv_{jn\ell}   ~&=&~ q^{k + (n-\ell-r_1)/2}\tv_{jn\ell} \ , \nn\\
\pi_{r_1,r_2}(K_2)\,\tv_{jn\ell}   ~&=&~ q^{\ell + (n-j-r_2)/2}\tv_{jn\ell} \ , \\
\pi_{r_1,r_2}(X^+_1)\,\tv_{jn\ell}   ~&=&~
 q^{-\ell} q_{12} \left({q_{12}q_{23}\over q_{13} }\right)^{(\ell+n)/2} [j+n-\ell-r_1]_q
\,\tv_{j+1,n\ell} ~+\nn\\ &&+~
 q^{j+n-\ell-r_1} q_{12}\left({q_{12}q_{23}\over q_{13} }\right)^{(\ell-n)/2 -1} [\ell]_q
 \,\tv_{j,n+1,\ell-1} \ ,\\
\pi_{r_1,r_2}(X^+_2)\,\tv_{jn\ell}   ~&=&~ - {q^{r_2-1-n} q_{13}\over q_{12}}
\left({q_{12}q_{23}\over q_{13} }\right)^{(j+\ell+n)/2} [j]_q
\,\tv_{j-1,n+1,\ell} ~-\nn\\
&&- \ q^{-j}  q_{23}
\left({q_{12}q_{23}\over q_{13} }\right)^{(j+n)/2} [\ell-r_2]_q
\,\tv_{j,n,\ell+1}\ ,\\
\pi_{r_1,r_2}(X^-_1)\,\tv_{jn\ell}   ~&=&~ - {q^{\ell+1} \over   q_{12}}
\left({q_{13} \over q_{12}q_{23}  }\right)^{(\ell+n)/2} [j]_q
\,\tv_{j-1,n,\ell} \ ,\\
\pi_{r_1,r_2}(X^-_2)\,\tv_{jn\ell}   ~&=&~ - q^{\ell}
\left({q_{13} \over q_{12}q_{23}  }\right)^{(j+\ell-n)/2} [n]_q
\,\tv_{j+1,n-1,\ell} \ - \\
&&- {q^{n+1}\over q_{23}}
\left({q_{13} \over q_{12}q_{23}  }\right)^{(j+n)/2} [\ell]_q
\,\tv_{j,n,\ell-1} \
\eena


Further, since the action of ~$\cu$~ is not affecting the degrees
of $\cd_i$,  we introduce (as in \cite{Dob})  the restricted functions
~$\hp ({\bar Y})$~ by the
formula which is prompted in \eqref{serasa}~:
\eqn{res} \hp({\bar Y}) ~\equiv~ \bigl( \hca\tv ) ({\bar Y}) ~\doteq ~\tv
({\bar Y}, \cd_1 = \cdots = \cd_{n-1} = {\bf 1}) ~. \ee
We denote the representation space of ~$\hp({\bar Y})$~ by
~$\tcc_{\br}$~ and the representation acting in ~$\tcc_{\br}$~
by ~$\hat\pi_{\br}$~. ~Thus, the operator ~$\hca$~ acts from
~$\cc_{\br}$~ to ~$\tcc_{\br}$~. ~The properties of
~$\tcc_{\br}$~ follow from the intertwining requirement for ~$\hca$~
\cite{Dob}:
\eqn{int} \hat\pi_{\br} ~\hca ~~=~~ \hca ~ \tilde
\pi_{\br} ~.\ee

Thus, the representations ~$\hat\pi_{\br}$~ have the same action on
~$\hp$~ as given for ~$\tv$~ in \eqref{ulpiz}.

Next we recall that we have defined the representations ~$\hat\pi_{\br}$~ for
~$r_i\in\bbz$. However, from \eqref{ulpiz} we notice that we
can consider the restricted functions $\hp({\bar Y})$ for
arbitrary complex $r_i$. We shall make this extension from now on.

\vskip 5mm

\section*{Outlook}

 We have constructed  representations  of the quantum
algebras ~$U_{q{\bf q}}(gl(n))$\\ and ~$U_{q{\bf q}}(sl(n))$. The
representations are labelled by $n-1$ complex numbers ~$r_i$~ and
are acting in the space of formal power series of ~$1+n(n-1)/2$~
non-commuting variables which  variables generate  quantum flag
manifolds  of ~$GL_{q\q}(n)$, ~$SL_{q\q}(n)$. The case $n=3$ was
treated explicitly. The next tasks in preparation \cite{Doprep} are
to find the reducibility conditions of our representations and to
  construct the quantum difference intertwining operators in this
  multiparameter setting.

\vskip 10mm

\section*{Acknowledgments} The author has received partial support from COST
 Action MP1405, from Bulgarian NSF Grant DN-18/1 and from PHC Rila.

\end{document}